\documentclass{aa}  
\usepackage{graphicx} 
\usepackage{txfonts}
\usepackage{amsmath}
\usepackage{amssymb}
\usepackage{lipsum}
\usepackage[colorlinks=true,linkcolor=blue]{hyperref}
\usepackage{caption}
\usepackage[table]{xcolor}
\usepackage{lineno}
\begin{document} 

\nolinenumbers

   \title{Galaxy cluster profiles: A Gaussian mixture model approach to halo miscentering}

   \author{Matthew Currie\inst{1}\thanks{curriem@sas.upenn.edu}, Kyle Miller\inst{1}\thanks{mky@sas.upenn.edu}, Tae-Hyeon Shin\inst{2}, Eric Baxter\inst{3} \and Bhuvnesh Jain\inst{1}
          }

   \institute{Department of Physics and Astronomy, University of Pennsylvania, Philadelphia, Pennsylvania, USA
         \and
             Department of Physics and Astronomy, Stony Brook University, Stony Brook, New York, USA
         \and
             Institute for Astronomy, University of Hawaii, Mānoa, Hawaii, USA
             }

   \date{Received XXX; accepted XXX}

\titlerunning{Mixture model for miscentering}
\authorrunning{M. Currie and K. Miller et al.}

 
  \abstract
   {Measurements of the galaxy density and weak-lensing profiles of galaxy clusters typically rely on an assumed cluster center, which is  taken to be the brightest cluster galaxy or other proxies for the true halo center.  Departure of the assumed cluster center from the true halo center bias the resultant profile measurements, an effect known as miscentering bias.}
   {Currently, miscentering is typically modeled in stacked profiles of clusters with a two parameter model. We use an alternate approach in which the profiles of individual clusters are used with the corresponding likelihood computed using a Gaussian mixture model.}
   {We test the approach using halos and the corresponding subhalo profiles from the IllustrisTNG hydrodynamic simulations.}
   {We obtain significantly improved estimates of the miscentering parameters for both 3D and projected 2D profiles relevant for imaging surveys. We discuss applications to upcoming cosmological surveys.\thanks{Our \textsc{Python} package for implementing the Gaussian mixture model is publicly avaliable at https://github.com/KyleMiller1/Halo-Miscentering-Mixture-Model}}
   {}


   \maketitle
%
\section{Introduction}
\label{sec:introduction}

    The matter and the galaxy density profiles of massive halos carry valuable information on the cosmic evolution and astrophysics. For example, estimation of the total mass distribution of observed galaxy clusters for cluster cosmology relies on accurate measurements of their weak-lensing profiles \citep{Mantz2015,Costanzi2019,Ghirardini2024,Bocquet2024}. Also, the galaxy number density profiles around massive clusters could be used to constrain e.g. the physical boundary of halos and the timescale for star formation of infall galaxies to be quenched \citep[e.g.][]{Shin2021,Adhikari2021}. Correctly measuring the profiles of the central region of halos ($\lesssim 1{\rm Mpc}$) is valuable, since it is highly sensitive to halo quantities such as the total halo mass \citep[e.g.][]{Melchior2017}, baryonic activities \citep[e.g. AGN feedback;][]{Peirani2017} and dynamical friction to subhalos \citep[e.g.][]{Adhikari2016}. Estimating the halo center is critical to obtain accurate profile.  Since the true center is unknown, the realistic goal is to understand how the observational proxy for the halo center deviates from the true center (miscentering), and retrieve the miscentering distribution of a given set of observed halos; failure to do so would result in biased density profiles and therefore biased results.

    The position of the center of a massive halo is estimated in several ways from observations; the position of the brightest central galaxy \citep[BCG; e.g.][]{Rykoff2014}, the local peak of the Sunyaev-Zel'dovich effect \citep{SZ1972} from the intracluster gas \citep{Bleem2015,Hilton2021}, or the peak of the extended X-ray emission \citep{Hollowood2019,Bulbul2024}. However, due to systematic effects such as halo mergers, projection effects and detector resolution limit, a certain fraction of these central positions could be misidentified or offset from the true halo center \citep{Kelly2023,Seppi2023,Roche2024}. While it is  possible to directly constrain the miscentering properties of observed halos using techniques such as comparing the BCG position to the X-ray peak \citep[which is proportional to the density squared, therefore less susceptible to sysetematics, e.g.][]{Zhang2019} and using realistic hydrodynamical simulations from which mock observations could be constructed \citep{Yan2020,Seppi2023}, it is  difficult to do this for the entire observed halo population due to the observational limits (e.g. time, redshift, resolution) and the uncertainty in the baryonic physics.

    For this reason, studies that involve fitting observational data typically characterize miscentering via free parameters in their models, namely the fraction of miscentered halos and the average amplitude of miscentering. Since individual halos from an observation are noisy (galaxy number density, weak-lensing, etc.), the analyses are  performed by stacking signals from the entire halo sample and fitting to a halo model \citep{McClintock2019,Shin2021} with an assumed distribution model of miscentering. 
    Any deviation in the posteriors of miscentering parameters of the stacked profiles from the truth would cause a bias in the final result. Furthermore, the stacking procedure, equivalent to taking an average of the profiles, involves information loss which contributes to a higher statistical uncertainty of the constrained halo profile. 
    
    To improve on the stacked profile approacing, we introduce a Gaussian mixture model to fit a collection of miscentered halo profiles to a halo model and estimate the miscentering parameters via Bayesian Markov Chain Monte Carlo (MCMC) inference. The Gaussian mixture model utilizes individual halo profiles to maximize the information usage. We assess the accuracy and the precision of the Gaussian mixture model relative to the stacking method by measuring subhalo number density profiles around the simulated halos from the IllustrisTNG simulation \citep{Nelson2019}. 

    The outline of this paper is as follows. In Sec.~\ref{sec:data-and-measure}, we describe the IllustrisTNG simulation data and the procedure to measure the subhalo number density profile around each of the selected halos. The halo model and the two approaches (stacked, Gaussian mixture) for the miscentering as well as the procedure for the MCMC fitting are described in Sec.~\ref{sec:method}. We show our main results in Sec.~\ref{sec:result} and conclude in Sec.~\ref{sec:conclusion}. We use the capital letter $R$ for 2-dimensional distances and the lowercase $r$ to denote 3-dimensional distances.

\section{Data and measurement}
\label{sec:data-and-measure}

    \subsection{The IllustrisTNG simulation}
    \label{sec:data_simulation}

        The IllustrisTNG simulation \citep{Nelson2019,Pillepich2018,Marinacci2018,Naiman2018,Springel2018} is a suite of magnetohydrodynamic simulation, built with an improved treatment of baryonic physics from the original Illustris simulation suite \citep{Vogelsberger2014a,Genel2014,Sijacki2015,Vogelsberger2014b} using the moving-mesh \textsc{AREPO} code. A flat $\Lambda$CDM cosmology is used with the parameter values $\Omega_{\rm m}=0.3089$, $\Omega_{\rm b}=0.0486$, $\sigma_8=0.8159$, $n_{\rm s}=0.9667$ and $H_0=67.74\, {\rm km \, s^{-1} Mpc^{-1}}$. 

        We use the IllustrisTNG 300-1 simulation, the highest resolution cube with the side length of 302.6 Mpc. It comprises an initial set of 2$\times$$2500^3$ resolution elements, with the dark matter mass resolution of 5.9$\times$$10^7$, the baryon mass resolution of 1.1$\times$$10^7$, the minimum gas softening length of 370 pc, and the dark matter softening length of 1.48 kpc. The public data release provides 99 snapshots between $z=20$ and $z=0$ and we use the snapshot at $z=0$. The halo catalog is constructed using the friends-of-friends group-finding algorithm with the subhalos identified with \textsc{Subfind} algorithm \citep{Springel2001,Dolag2009}.

\begin{figure}
\includegraphics[width=0.99\columnwidth]{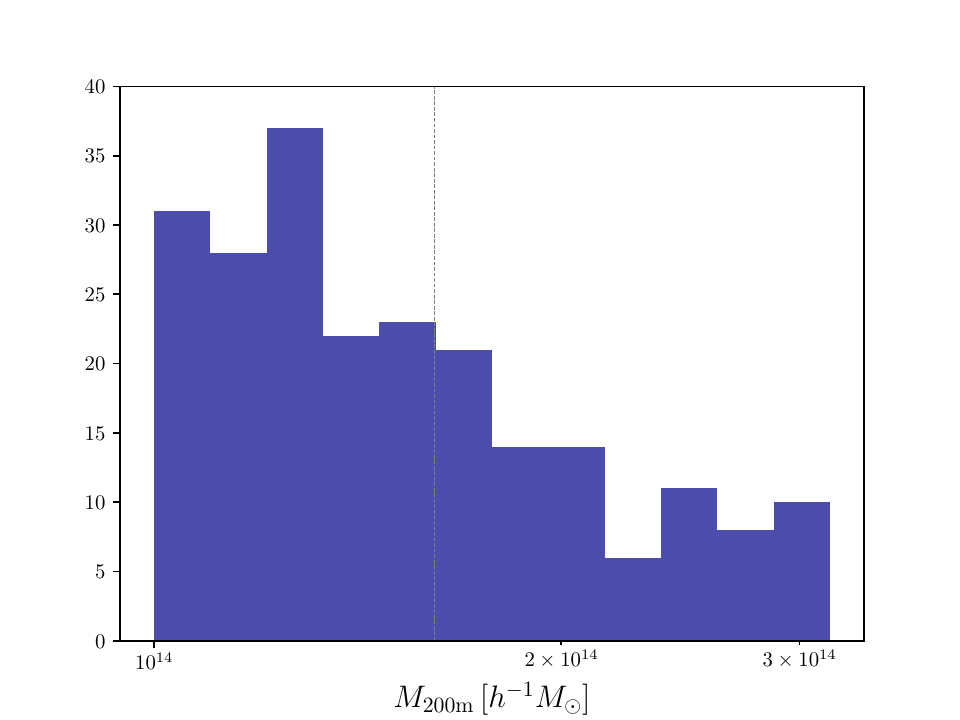}
\caption{The mass distribution of the 225 halos from the simulation. The vertical grey line represents the mean mass of the sample.} 
\label{fig:halo_mass}
\end{figure}

        To select massive halos of galaxy cluster sizes, we consider parent halos within the mass range of $M_{\rm 200m}=[10^{14},10^{14.5}] \, h^{-1} M_\odot$, where $M_{\rm 200m}$ represents the mass enclosed within the radius where the mean mass density inside becomes 200 times the cosmic mean matter density. The center of each halo is determined by identifying the gravitationally most bound particle. To calculate the subhalo number density profiles around the selected parent halos, we take subhalos with stellar mass larger than $10^8 \, h^{-1} M_{\odot}$ and with a non-zero \texttt{SubhaloFlag} to exclude those of non-cosmological origin. The final parent halo sample includes 225 halos with the mean mass of $M_{\rm 200m}=1.61$$\times$$10^{14} \, h^{-1} M_\odot$. We show the distribution of their masses in Fig.~\ref{fig:halo_mass}.

    \subsection{The subhalo density profiles}
    \label{sec:measurement_profiles}

        From the simulation halo catalog, we randomly select a fraction, $f_{\rm mis}$, of halos to be miscentered and then apply offset values chosen from a Rayleigh distribution \citep[Eq.~\ref{eq:rayleigh},][]{Rykoff2016} with a scale parameter of $\sigma_{\rm R}$. Motivated by observations, we test the miscentering models with two values per each parameter: $f_{\rm mis}\in\{0.22,0.50\}$ and $\sigma_{\rm R}\in\{0.3,0.5\} R_{\rm 200m}$. The miscentering direction is also randomly chosen per halo. To boost our signal while maintaining a quasi-independent halo sample, we make three realizations of the miscentered profiles per halo with independently chosen miscentering properties as above, resulting in 675 halos in our final sample.

        We use the subhalo number density profile of each halo as a proxy for the total mass density profile, which is computed in 14 logarithmically spaced radial bins between $0.1$ and $1.08 \, R_{\rm 200m}$. We normalize the distance in units of  $R_{\rm 200m}$, which is a natural choice for dark matter halos under $\Lambda$CDM cosmology. The covariance matrix, $ \mathbb{C}_i$, for the $i$-th halo profile is computed using the Jackknife resampling method with 70 other halos closest in mass to the given halo, whereas the covariance for the stacked measurement is computed with the Jackknife method using all halos. For the simplicity of the analysis, we only use the diagonal components of the covariance matrix. While we note that this analysis choice for the covariance estimation deviates from the real observational cases, we argue that it suffices our purpose of comparing the two miscentering models (the stacked model, the Gaussian mixture model). 

        However, observational data typically do not come with perfect information of the 3-dimensional distribution of halos. To  mimic this situation, we also construct the 2-dimensional subhalo number density profiles for each parent halo, by projecting the 3-dimensional halo and subhalo positions onto a 2-dimensional plane. Each of the halos and its two artificial copies (see above) are projected onto a separate plane, chosen among the $xy$, $yz$, and $xz$-planes. We miscenter these 2-dimensional profiles similarly as above, but with the miscentering directions being on the 2-dimensional plane. The 2-dimensional number density profiles are calculated using 10 logarithmically spaced radial bins between $0.1$ and $1.08 \, R_{\rm 200m}$. To extract information from a given parent halo, we subtract the mean surface number density of the subhalos over the entire simulation box from the measured 2-dimensional profiles. The covariance matrix for each profile is computed similarly as done in the 3-dimensional case, with the 70 other halos closest in mass and having the same projection axis.

\section{Methodology}
\label{sec:method}

    \subsection{The halo model}
    \label{sec:halo_model}

        In this section, we describe the halo model we use to fit the subhalo density profiles measured as described in Sec.~\ref{sec:measurement_profiles}. We adopt the model proposed in \citet{Diemer2023} (D23). This model consists of the term that describes the matter orbiting around a halo (orbiting term) and the term that describes the matter falling into the halo. However, since we consider subhalo density profiles only up to $\sim$$R_{\rm 200m}$, where the orbiting matter dominates and the effect of miscentering is the most visible, we only keep the orbiting term. We refer readers to \citet{Diemer2023} for details of this model and briefly describe our choice of halo model here.

        The orbiting term, $\rho_{\rm orbit}$, is defined as 
        \begin{equation}
        \label{eq:rho_orbit}
        \rho_{\text{orbit}}(r) = \rho_{\text{s}} e^{S(r)},
        \end{equation}
        where
        \begin{equation}
        \label{eq:Sr}
        S(r) = \frac{-2}{\alpha} \left[ \left(\frac{r}{r_{\text{s}}}\right)^{\alpha} - 1 \right] - \frac{1}{\beta} \left[ \left(\frac{r}{r_{\text{t}}}\right)^{\beta} - \left(\frac{r_{\text{s}}}{r_{\text{t}}}\right)^{\beta} \right].
        \end{equation}
        Note that the first term in $S(r)$ is the Einasto profile with slope parameter, $\alpha$, and a scale radius, $r_{\rm s}$. The second term in $S(r)$ truncates the orbiting term at large distances as one approaches the infall region. This model is proven to be able to accurately fit the dark matter halos in N-body simulations \citep{Diemer2023}. Here, $r_{\rm t}$ is the truncation radius and $\beta$ modulates the steepness of the truncation. $\rho_{\rm s}$ normalizes the profile. Therefore, in this model, we have five free parameters, $\{\rho_{\rm s}, r_{\rm s}, \alpha, r_{\rm t}, \beta\}$.

        The 2D projected halo model is obtained by integrating the 3D model above along the line-of-sight direction:
        \begin{equation}
            \Sigma_{\text{orbit}}(R) = \int_{-l_{\rm max}}^{l_{\rm max}} {{\rm d} l \, \rho_{\rm orbit}\left(\sqrt{R^2 + l^2}\,\right)} \, ,
        \label{eq:6}
        \end{equation} 
        where $R$ is the projected halo-centric distance, and we take $l_{\rm max} = 40 \, R_{\rm 200m}$, the maximum line-of-sight distance of integration. We have checked that increasing the maximum integration scale does not meaningfully affect our result.

    \subsection{The stacked model}
    \label{sec:stacked_model}

        In this section, we describe the miscentering model that fits stacked (averaged) profile data, which is a commonly used approach in many studies. 
        Under the assumption of an isotropic profile, the surface density profile of a halo miscentered by a distance $R_{\rm mis}$, $\Sigma_{\rm mis}(R|R_{\rm mis})$, is expressed as \citep{Shin2019,McClintock2019}:
        \begin{equation}
        \label{eq:Sigma_RRmis}
        \Sigma_{\text{mis}}(R | R_{\text{mis}}) = \frac{1}{2 \pi}\int_{0}^{2 \pi} {{\rm d}\phi \, \Sigma\left(\sqrt{R^2 + R_{\text{mis}}^2 + 2 R R_{\text{mis}} \cos{\phi}}\,\right)},
        \end{equation}
        where the integral is performed over the azimuthal angle, $\phi$, and $\Sigma$ is the surface density profile without miscentering. 

        For a  distribution of $R_{\rm mis}$ given by $P(R_{\rm mis})$, the stacked surface density profile with the given miscentering distribution, $\Sigma_{\rm mis}$, is given by:
        \begin{equation}
        \label{eq:Sigma_mis}
            \Sigma_{\rm mis}(R) = \int {{\rm d}R_{\rm mis} \, P(R_{\rm mis}) \, \Sigma_{\rm mis}(R | R_{\rm mis})}.
        \end{equation}
        As in \citet{Shin2019}, we assume a Rayleigh distribution for the miscentering distribution of halos, $P(R_{\rm mis})$:
        \begin{equation}
        \label{eq:rayleigh}
            P(R_{\rm mis}) = \frac{R_{\rm mis}}{\sigma_{\rm R}^2} \exp\left[{\frac{-R_{\rm mis}^2}{2\sigma_{\rm R}^2}}\right],
        \end{equation}
        which accurately describes the miscentering of observed galaxy clusters \citep{Rykoff2016}. The factor $\sigma_{\rm R}$ modulates the  miscentering amplitudes of the halos. 
        
        Finally, if a fraction, $f_{\rm mis}$, of the stacked halos are miscentered according to the above distribution while the others are correctly centered, the stacked surface density profile is:
        \begin{equation}
        \Sigma(R) = (1 - f_{\rm mis}) \Sigma_0(R) + f_{\rm mis} \Sigma_{\rm mis}(R),
        \end{equation} 
        where $\Sigma_0$ is the correctly centered profile. 
        
        In the case of the 3-dimensional analysis, the expression for a halo profile miscentered by $r_{\rm mis}$, $\rho_{\rm mis}(r|r_{\rm mis})$, could be analogously derived by taking a spherical average. Since we assume an isotropic halo density profile, the azimuthal dependence could be handled by symmetry, leaving us with:
        \begin{equation}
        \label{eq:rho_rrmis}
            \rho_{\rm mis}(r | r_{\rm mis}) = \frac{1}{2}\int_{0}^{\pi} {{\rm d}\vartheta \, \sin{\vartheta} \, \rho\left(\sqrt{r^2 + r_{\rm mis}^2 + 2 r r_{\rm mis} \cos{\vartheta}}\,\right)},
        \end{equation} 
        where $\vartheta$ represents the polar angle and $\rho$ is the correctly centered profile.
        
        Then a stacked density profile in 3D following a Rayleigh miscentering distribution could be expressed analogously as in 2D as:
        \begin{equation}
        \rho_{\rm mis}(r) = \int {{\rm d}r_{\rm mis} \, P(r_{\rm mis}) \, \rho_{\rm mis}(r | r_{\rm mis})}
        \end{equation}
        and
        \begin{equation}
        \rho(r) = (1 - f_{\rm mis}) \rho_0(r) + f_{\rm mis} \rho_{\rm mis}(r),
        \end{equation}
        where $\rho_0$ is the correctly centered profile.
        
        To fit the data to the model, we assume a Gaussian likelihood for the halo profile:
        \begin{equation}
        \mathcal{L}(\vec{d} | \vec{\theta}) \propto \exp \left[ -\frac{1}{2} \left(\vec{d} - \vec{m}(\vec{\theta})\right)^T \mathbb{C}^{-1}\left(\vec{d} - \vec{m}(\vec{\theta})\right) \right],
        \end{equation}
        where $\vec{d}$ is the stacked profile measurement from data, $\vec{m}(\vec{\theta})$ is the model for the miscentered profile described above, and $\mathbb{C}$ is the covariance matrix. Here, $\vec{\theta}$ is the set of free parameters in our model, $\{\rho_{\rm s}, r_{\rm s}, \alpha, r_{\rm t}, \beta, f_{\rm mis}, \sigma_{\rm R}\}$ (five halo parameters and two miscentering parameters). The priors for the seven model parameters are described in Table~\ref{tab:1}. Note that we use uninformative uniform priors for the miscentering parameters to  gauge the model's ability to retrieve the true values.
        
\begin{table}
	\centering
	\resizebox{\columnwidth}{!}{\begin{tabular}{lll}
		Parameter & Prior & Description \\ \hline
		$\log \rho_{\rm s}$ & $[-20, 20]$ & overdensity at scale radius \\
		$\log \, (r_{\rm s} / R_{\rm 200m})$ & [$\log (0.01)$, $\log (0.45)$] & scale radius \\
        $\log \, (r_{\rm t} / R_{\rm 200m})$ & [$\log (0.5)$, $\log (10)$] & truncation radius \\
        $\log \alpha$ & [$\log (0.03)$, $\log (0.4)$] & radial evolution of slope \\
		$\log \beta$ & [$\log (0.1)$, $\log (10)$] & sharpness of truncation \\
        $f_{\rm mis}$ & $[0, 1]$ & miscentering fraction \\
        $\sigma_{\rm R}$ & $(0, 1]$ & miscentering amplitude \\
        \hline
	\end{tabular}}
    \caption{The prior range of each model parameter used in this study. The priors for the five halo parameters are based on the recommendation in \citet{Diemer2023}.}
    \label{tab:1}
\end{table}

    \subsection{The Gaussian mixture model}
    \label{sec:mixture-model}

        The Gaussian mixture model uses likelihoods for individual halo profiles, which are then combined into one:
    \begin{equation}
        \label{eq:mixture_model}
            \mathcal{L}(\{ \vec{d}_i \} | \vec{\theta}) = \prod_i \mathcal{L}_i(\vec{d}_i | \vec{\theta}),
        \end{equation} 
        where $\vec{d}_i$ represents the measured halo profile for the $i$-th halo and $\vec{\theta}$ is the set of free parameters described in the previous section.
        
        The likelihood of the $i$-th profile, $\mathcal{L}_i(\vec{d}_i|\vec{\theta})$, is decomposed into the correctly centered term and the miscentered term:
        \begin{equation}
        \label{eq:mixture_model_individual}
        \begin{split}
        \mathcal{L}_i &(\vec{d}_i | \vec{\theta}) =  \,(1-f_{\rm mis})\mathcal{N}(\mu = \vec{d}_{\rm i} - \vec{m}_{\rm cen}(\vec{\theta}), \mathbb{C}_i ) \\
        &+ f_{\rm mis} \int {\rm d}R_{\rm mis}^i P(R^i_{\rm mis} \, | \, \sigma_{\rm R}) \mathcal{N}(\mu = \vec{d}_i - \vec{m}_{\rm mis}(\vec{\theta} \, | \, R^i_{\rm mis}), \mathbb{C}_i ),
        \end{split}
        \end{equation}
        where $\mathcal{N}$ is the Gaussian distribution, $\vec{m}_{\rm cen}$ is the correctly centered halo model, $\mathbb{C}_i$ is the covariance matrix of the $i$-th halo and $\vec{m}_{\rm mis}$ is the miscentered halo model in the previous section. Note that in this mixture model, the miscentering parameters ($f_{\rm mis}$, $\sigma_{\rm R}$) are explicitly expressed in the likelihood, rather than included in $\vec{\theta}$, and $\vec{m}_{\rm mis}(\vec{\theta} \, | \, R^i_{\rm mis})$ is the miscentered profile for an individual halo (Eq.~\ref{eq:Sigma_RRmis} for 2D and Eq.~\ref{eq:rho_rrmis} for 3D). We use the same priors (Table \ref{tab:1}) as the stacked model. Also note that while we use the information from the individual halo profile measurements for the mixture model, these individual profiles are still modeled with one universal set of halo parameters, included in $\vec{\theta}$.

\section{Result}
\label{sec:result}

    \subsection{Model fitting}
    \label{sec:result_fitting}

\begin{figure}
\includegraphics[width=0.99\linewidth]{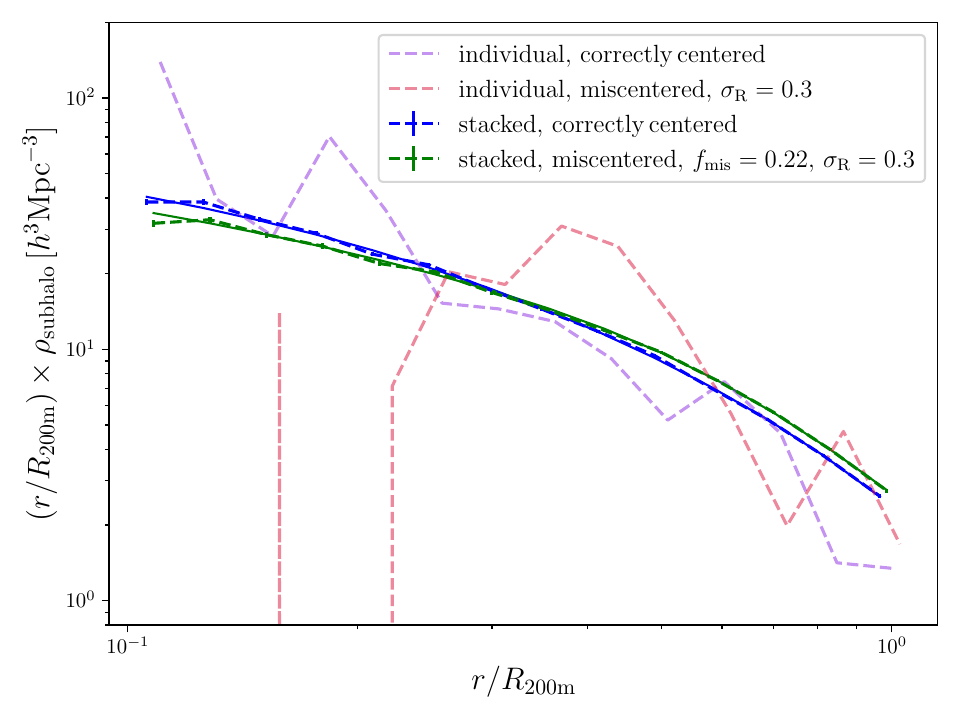}
\caption{
The measured 3D subhalo density profiles (Sec.~\ref{sec:measurement_profiles}) for the stacked profile with no miscentering (blue dashed), stacked profile with $f_{\rm mis}=0.22$ and $\sigma_{\rm R}=0.3$ (green dashed), an example individual profile with no miscentering (violet dashed), and an example individual profile with $\sigma_{\rm R}=0.3$ (red dashed). The solid curves with the corresponding colors are the best-fit models (see Sec.~\ref{sec:result_fitting}).
} 
\label{fig:profile_fit_3d}
\end{figure}

\begin{figure}
\includegraphics[width=0.99\linewidth]{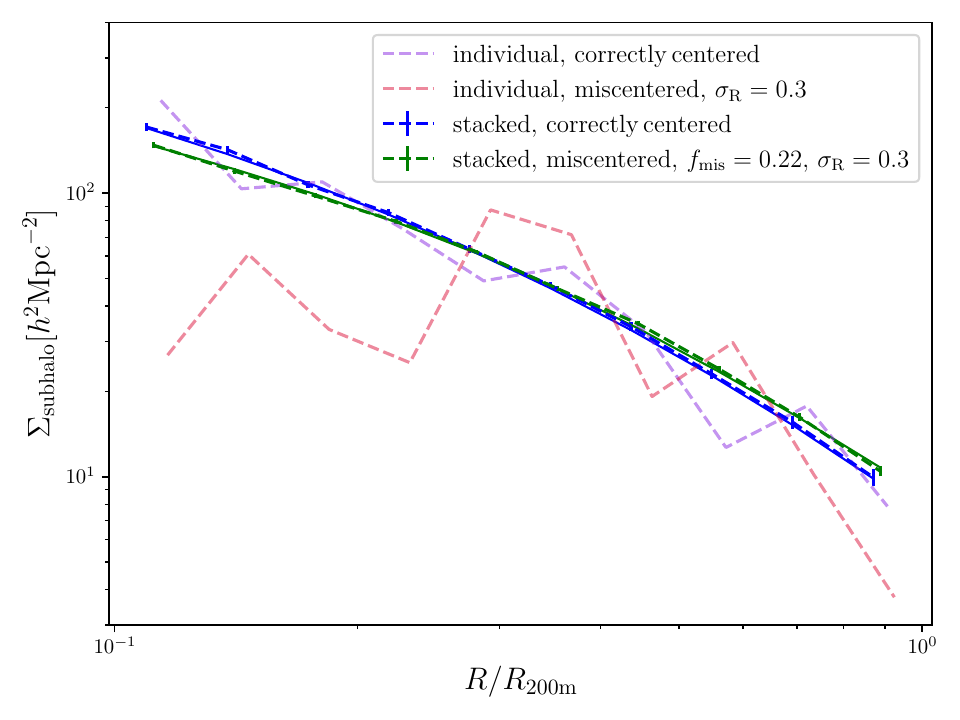}
\caption{Similar to  Fig~\ref{fig:profile_fit_3d}, but with the 2D measurements.
} 
\label{fig:profile_fit_2d}
\end{figure}

        The model fits were performed with \texttt{PyMultiNest} \citep{Buchner2014}, a Python interface for \texttt{MultiNest}, a nested sampling Markov Chain Monte Carlo algorithm made for Bayesian inference. For our analysis, we use a sampling efficiency of 0.8 and evidence tolerance of $10^{-2}$, where convergence of chains is assured by an internal algorithm of the package. We refer readers to the above reference for details of this algorithm.

        The Einasto slope parameter, $\alpha$, determines how rapidly the orbiting term (Eq.~\ref{eq:rho_orbit} and \ref{eq:Sr}) steepens, suggesting a degenerate relationship with the miscentering parameters. Accordingly, we consider the model performance both in the case where $\alpha$ is free (Table \ref{tab:1}) and informed in prior. For the latter, we separately perform a model fit (sec.~\ref{sec:method}) without having any miscentering effect in our data and model, then use the resultant constraint on $\alpha$ as a prior for the main analysis. This preliminary fitting returned $\log(\alpha) = -0.524 \pm 0.074$ for the 3D analysis and $\log(\alpha) = -0.707 \pm 0.104$ for the 2D analysis.
 
        In Fig.~\ref{fig:profile_fit_3d}, we show the measurements of the 3D subhalo density profiles (Sec.~\ref{sec:measurement_profiles}) and their best-fit curves. We show profiles with no miscentering, and profiles with $f_{\rm mis}=0.22$ and $\sigma_{\rm R}=0.30 \, R_{\rm 200m}$ as a representative case. The other combinations of the miscentering parameters result in similar quality of fitting.

        The stacked subhalo density profile measurement around the correctly centered (miscentered) halos is plotted in dashed blue (green). The best-fit model for the measurement (Sec.~\ref{sec:stacked_model}) is shown via solid curves with the corresponding color. We also show an example profile of a randomly chosen individual halo with miscentering ($\sigma_{\rm R}=0.3$) in red and that without miscentering in violet. When fitting the correctly centered profiles, we do not include miscentering parameters in our model. 

        In Fig.~\ref{fig:profile_fit_2d}, we show a similar set of profile measurements and best-fit models for the 2D analysis with the same color scheme and configuration as Fig.~\ref{fig:profile_fit_3d}. Note that in this case miscentering is applied in 2D ($R_{\rm mis}$ rather than $r_{\rm mis}$). 

        For the stacked measurements, we use a $\chi^2$ tests to assess the quality of fitting, defined as 
        \begin{equation}
            \chi^2 = \left(\vec{d} - \vec{m}(\vec{\theta}_{\rm MAP})\right)^T \mathbb{C}^{-1} \left(\vec{d} - \vec{m}(\vec{\theta}_{\rm MAP})\right) \, ,
            \label{eq:chisq}
        \end{equation} 
        where $\vec{\theta}_{\rm MAP}$ is the set of maximum a posteriori (MAP) model parameters from the MCMC chains, and $\vec{d}$ and $\mathbb{C}$ are the stacked profile measurement from data and the covariance matrix (Sec.~\ref{sec:measurement_profiles}), respectively. Due to the complexity of the likelihood for the mixture model (Eq.~\ref{eq:mixture_model}, \ref{eq:mixture_model_individual}), we do not report probabilities-to-exceed (PTEs) for the mixture model fitting. We have checked that the mixture model gives reasonable fitting to the individual halo profiles, while the validity of using D23 model to simulated halos is rigorously verified in \citet{Diemer2023}. 
        
\begin{table}
\centering
\begin{tabular}{c|c|c}
    \textbf{PTE (stacked model)} & \textbf{3D} & \textbf{2D} \\
    \hline
    \hline
    No miscentering & 0.02 & 0.35 \\
    \hline
    $f_{\rm mis} = 0.22$, $\sigma_{\rm R} = 0.30$ & 0.02 & 0.10 \\
    $f_{\rm mis} = 0.22$, $\sigma_{\rm R} = 0.50$ & 0.08 & 0.08 \\
    \hline
    $f_{\rm mis} = 0.50$, $\sigma_{\rm R} = 0.30$ & 0.04 & 0.05 \\
    $f_{\rm mis} = 0.50$, $\sigma_{\rm R} = 0.50$ & 0.36 & 0.44 \\
    \hline
\end{tabular}
\caption{The probability-to-exceed (PTE) values from the model fitting to the stacked measurements with different miscentering configurations.}
\label{tab:PTE}
\end{table}

        We calculate PTEs of the best-fit models from a $\chi^2$ distribution with the degrees of freedom, $\nu = N_R - N_F$, where $N_R$ is the number of radial bins (14 for 3D, 10 for 2D) and $N_F$ is the number of free parameters in the model (5 for D23 model, 2 for miscentering). The PTEs of the fits for all miscentering parameter combinations are summarized in Table~\ref{tab:PTE} showing reasonable quality of fitting, although a few 3D cases exhibit marginally acceptable values (0.02-0.04). From Fig.~\ref{fig:profile_fit_3d} (blue and green curves), we can see these small PTE values are driven by  the slight bin-to-bin data fluctuations, coupled with the small estimated errors -- suggesting that the covariance matrix may be slightly underestimated.

    \subsection{Constraints on miscentering parameters from 3D profiles}
    \label{sec:result_3D}

\begin{figure}
\includegraphics[width=0.99\linewidth]{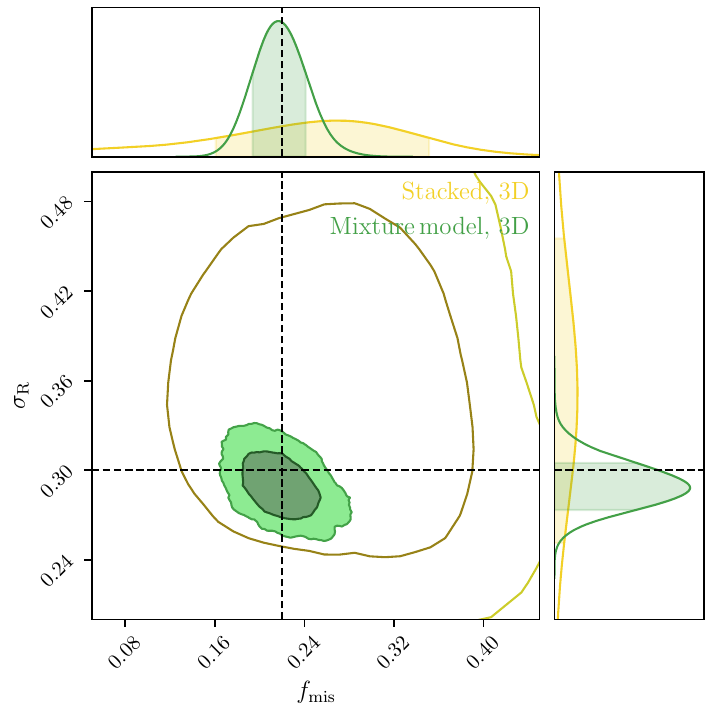}
\caption{The posterior contours from the 3D analysis with $f_{\rm mis}=0.22$ and $\sigma_{\rm R}=0.30 \, R_{\rm 200m}$, with $\alpha$ (halo inner slope parameter) free. The contours obtained from the mixture model are far more constraining than the stacked model.  
\newline \textbf{Mixture Model}: $f_{\rm mis} = 0.22^{+0.02}_{-0.02}$, $\sigma_{\rm R} = 0.29^{+0.02}_{-0.01}$. 
\newline \textbf{Stacked Model}: $f_{\rm mis} = 0.25^{+0.07}_{-0.11}$, $\sigma_{\rm R} = 0.36^{+0.08}_{-0.07}$.} 
\label{fig:contour_3D}
\end{figure}

        We compare the performance of the two models (stacked model, mixture model, Sec.~\ref{sec:method}) in terms of how well each model retrieves the true miscentering parameters in the 3D profile analyses. Since the inner halo slope parameter $\alpha$ is degenerate with the miscentering parameters, we also consider the case where we use an informative prior for $\alpha$ (see Sec.~\ref{sec:result_fitting} for the prior ranges used). As discussed in Sec.~\ref{sec:measurement_profiles}, we test our models with two values of each miscentering parameters, $f_{\rm mis}=\{0.22, 0.50\}$ and $\sigma_{\rm R}=\{0.30, 0.50\} \, R_{\rm 200m}$, which represent typical miscentering \citep{Rykoff2016} and large miscentering respectively. When presenting the results, the other five halo parameters are marginalized over in order to only extract information on miscentering.
        
        In Fig.~\ref{fig:contour_3D}, the posterior contours and the corresponding 1D marginalized distributions of the two miscentering parameters are shown (stacked: yellow, mixture: green), for the case with $f_{\rm mis}=0.22$, $\sigma_{\rm R}=0.3 R_{\rm 200m}$ and the flat prior for $\alpha$ as a representative example.
        
        While both models retrieve the true values of miscentering parameters within 1-$\sigma$, the mixture model (green) exhibits a significantly tighter constraint,  by a factor of $\sim$4-5, than the stacked model (yellow). 

\begin{table*}
	\centering
	\begin{tabular}{l||l|l||l|l}
		\textbf{True value} & \textbf{Mixture model} & \textbf{Mixture model} ($\alpha$ prior) & \textbf{Stacked model} & \textbf{Stacked model} ($\alpha$ prior) \\ \hline \hline 
        $f_{\rm mis} = \textbf{0.22}$ & $0.22^{+0.02}_{-0.02}$ & $0.22^{+0.02}_{-0.02}$ & $0.25^{+0.07}_{-0.11}$ & $0.25^{+0.08}_{-0.12}$
        \\ 
        $\sigma_{\rm R} = \textbf{0.30}$ & $0.29^{+0.02}_{-0.01}$ & $0.29^{+0.01}_{-0.01}$ & $0.36^{+0.08}_{-0.07}$ & $0.34^{+0.10}_{-0.06}$ \\ \hline 

        $f_{\rm mis} = \textbf{0.22}$ & $0.23^{+0.01}_{-0.02}$ & $0.23^{+0.01}_{-0.02}$ & $0.32^{+0.11}_{-0.14}$ & $0.33^{+0.07}_{-0.10}$
        \\ 
        $\sigma_{\rm R} = \textbf{0.50}$ & $0.51^{+0.02}_{-0.02}$ & $0.51^{+0.02}_{-0.02}$ & $0.47^{+0.09}_{-0.11}$ & $0.47^{+0.06}_{-0.06}$ \\ \hline \hline

        $f_{\rm mis} = \textbf{0.50}$ & $0.51^{+0.02}_{-0.03}$ & $0.51^{+0.02}_{-0.03}$ & $0.55^{+0.10}_{-0.08}$ & $0.53^{+0.05}_{-0.06}$ \\ 
        $\sigma_{\rm R} = \textbf{0.30}$ & $0.30^{+0.01}_{-0.01}$ & $0.30^{+0.01}_{-0.01}$ & $0.31^{+0.02}_{-0.03}$ & $0.31^{+0.02}_{-0.02}$ \\ \hline 

        $f_{\rm mis} = \textbf{0.50}$ & $0.50^{+0.02}_{-0.02}$ & $0.50^{+0.02}_{-0.02}$ & $0.50^{+0.08}_{-0.09}$ & $0.48^{+0.08}_{-0.10}$
        \\ 
        $\sigma_{\rm R} = \textbf{0.50}$ & $0.51^{+0.02}_{-0.01}$ & $0.51^{+0.02}_{-0.01}$ & $0.39^{+0.08}_{-0.05}$ & $0.45^{+0.05}_{-0.05}$ \\ \hline \hline 
        
	\end{tabular}
    \caption{The 1D posteriors for $f_{\rm mis}$ and $\sigma_{\rm R}$ from fitting the 3D subhalo density profiles. The first two columns show the result for the mixture model and the last two columns the result for the stacked model. The left panels show the result with the flat $\alpha$ prior and the right panel shows the result with the informative prior for $\alpha$ (Sec.~\ref{sec:result_fitting}). $\sigma_{\rm R}$ is expressed in units of $R_{\rm 200m}$.}
    \label{tab:posteior_all_3D} 
\end{table*}

        In Table~\ref{tab:posteior_all_3D}, we show posterior ranges of the miscentering parameters for all 4 miscentering combinations and 2 miscentering models, with and without the informative prior on $\alpha$ (Sec.~\ref{sec:result_fitting}). The left two columns correspond to the mixture model and the right two columns to the stacked model. 
        In all cases, both models successfully retrieve the truth values within about 1-$\sigma$, while the mixture model exhibits much higher precision than the stacked model as the example above.
        We also note that including the informative prior for $\alpha$ (profile slope) does not change the performance of the models significantly.

    \subsection{Constraints on miscentering parameters from 2D profiles}
    \label{sec:result_2D}

\begin{figure}
\includegraphics[width=0.99\linewidth]{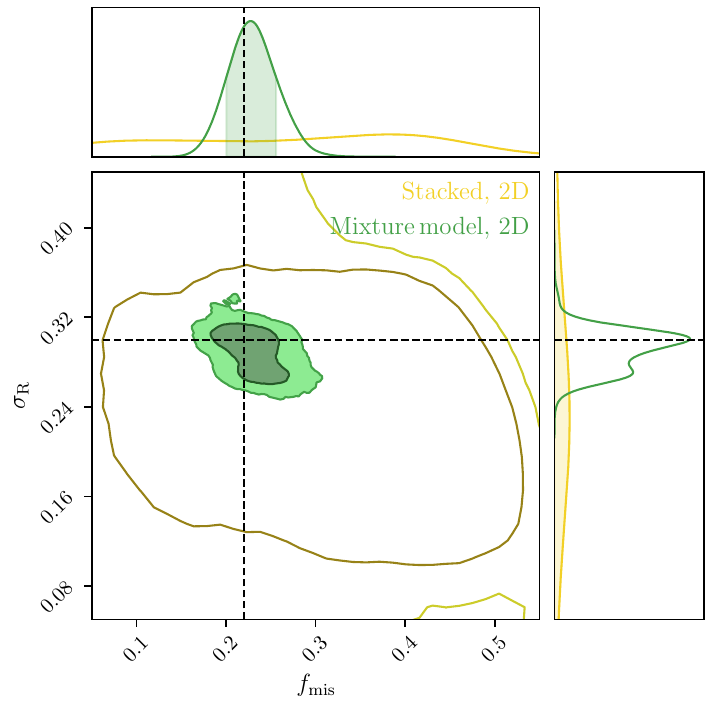}
\caption{A similar posterior contour plot as Fig.~\ref{fig:contour_3D}, but for the 2D analysis. The mixture model performs much better than the stacked model for this case as well. 
\newline \textbf{Mixture Model}: $f_{\rm mis} = 0.23^{+0.03}_{-0.03}$, $\sigma_{\rm R} = 0.29^{+0.02}_{-0.02}$. 
\newline \textbf{Stacked Model}: $f_{\rm mis} = 0.29^{+0.14}_{-0.20}$, $\sigma_{\rm R} = 0.24^{+0.46}_{-0.04}$.} 
\label{fig:contour_2D}
\end{figure}

\begin{table*}
	\centering
	\begin{tabular}{l||l|l||l|l}
		\textbf{True value} & \textbf{Mixture model} & \textbf{Mixture model} ($\alpha$ prior) & \textbf{Stacked model} & \textbf{Stacked model} ($\alpha$ prior) \\ \hline \hline 
        $f_{\rm mis} = \textbf{0.22}$ & $0.23^{+0.03}_{-0.03}$ & $0.23^{+0.03}_{-0.03}$ & $0.29^{+0.14}_{-0.20}$ & $0.26^{+0.08}_{-0.11}$
        \\ 
        $\sigma_{\rm R} = \textbf{0.30}$ & $0.29^{+0.02}_{-0.02}$ & $0.29^{+0.02}_{-0.01}$ & $0.24^{+0.46}_{-0.04}$ & $0.23^{+0.05}_{-0.05}$ \\ \hline 

        $f_{\rm mis} = \textbf{0.22}$ & $0.22^{+0.02}_{-0.02}$ & $0.22^{+0.01}_{-0.02}$ & $0.25^{+0.13}_{-0.16}$ & $0.13^{+0.19}_{-0.09}$
        \\ 
        $\sigma_{\rm R} = \textbf{0.50}$ & $0.56^{+0.04}_{-0.03}$ & $0.57^{+0.01}_{-0.02}$ & $0.21^{+0.62}_{-0.12}$ & $0.39^{+0.45}_{-0.35}$ \\ \hline \hline

        $f_{\rm mis} = \textbf{0.50}$ & $0.47^{+0.03}_{-0.04}$ & $0.47^{+0.03}_{-0.03}$ & $0.54^{+0.05}_{-0.10}$ & $0.48^{+0.05}_{-0.08}$ 
        \\ 
        $\sigma_{\rm R} = \textbf{0.30}$ & $0.30^{+0.02}_{-0.01}$ & $0.31^{+0.02}_{-0.01}$ & $0.29^{+0.03}_{-0.02}$ & $0.30^{+0.03}_{-0.04}$ \\ \hline 

        $f_{\rm mis} = \textbf{0.50}$ & $0.46^{+0.03}_{-0.03}$ & $0.46^{+0.03}_{-0.02}$ & $0.39^{+0.07}_{-0.09}$ & $0.34^{+0.11}_{-0.10}$
        \\ 
        $\sigma_{\rm R} = \textbf{0.50}$ & $0.54^{+0.04}_{-0.03}$ & $0.55^{+0.02}_{-0.03}$ & $0.35^{+0.15}_{-0.06}$ & $0.47^{+0.11}_{-0.10}$ \\ \hline \hline 

	\end{tabular}
    \caption{Similar as Table~\ref{tab:posteior_all_3D}, but using the 2D subhalo density profiles instead of 3D profiles.}
    \label{tab:posteior_all_2D}
\end{table*}

        From observational data  one generally has to reconstruct the 3D density profiles of observed halos from the projected profiles. So we also perform our analyses on the 2D projected subhalo density profiles measured as described in Sec.~\ref{sec:measurement_profiles}. 

        The configuration of the analyses is identical to the 3D case (Sec.~\ref{sec:result_3D}), except that we use 2D subhalo density profiles instead of the 3D profiles: 4 combinations of the applied miscentering parameters ($f_{\rm mis}=\{0.22,0.50\}$, $\sigma_{\rm R}=\{0.30, 0.50\} \, R_{\rm 200m}$), using two different miscentering model (stacked model, mixture model), with and without the prior in $\alpha$ (Sec.~\ref{sec:result_fitting}). 

        In Fig.~\ref{fig:contour_2D}, we plot the posterior contours from the model fitting with the true miscentering values of $f_{\rm mis}=0.22$ and $\sigma_{\rm R}=0.30 \, R_{\rm 200m}$ with a flat prior on $\alpha$. We first notice that, as in the 3D case, the mixture model results in significantly higher constraining power than the stacked model. 

        In Table~\ref{tab:posteior_all_2D}, we list the marginalized posterior ranges for the miscentering parameters for all analysis cases,  analogous to Table~\ref{tab:posteior_all_3D}. We confirm, as in the 3D case, the mixture model exhibits significantly higher precision than the stacked model. Furthermore, it is interesting that while the constraining power of the mixture model does not change much between 3D and 2D analyses, that of the stacked model becomes much weaker in many cases. 

    \subsection{Effects of informative miscentering priors}
    \label{result_informed_miscen}

\begin{figure}
\includegraphics[width=0.99\linewidth]{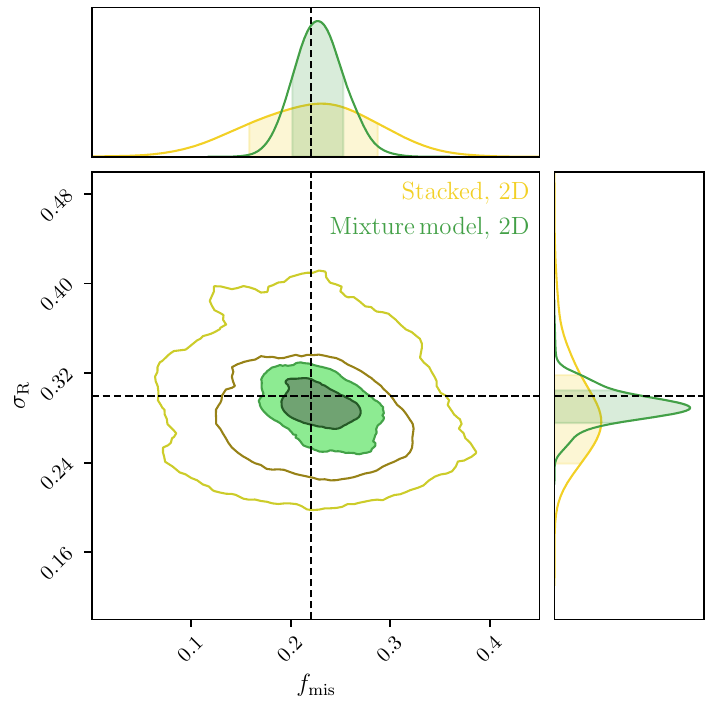}
\caption{A similar posterior contour plot as Fig.~\ref{fig:contour_2D}, but with priors on two miscentering parameters ($f_{\rm mis}=0.22\pm0.11$, $\sigma_{\rm R}=0.30\pm0.22$). The stacked model performs better with the priors but is still weaker than the mixture model. 
\newline \textbf{Mixture Model}: $f_{\rm mis} = 0.23^{+0.02}_{-0.02}$, $\sigma_{\rm R} = 0.29^{+0.02}_{-0.01}$. 
\newline \textbf{Stacked Model}: $f_{\rm mis} = 0.22^{+0.06}_{-0.06}$, $\sigma_{\rm R} = 0.28^{+0.04}_{-0.05}$.} 
\label{fig:contour_2D_miscenprior}
\end{figure}

        Finally, we test our models assuming that we have prior knowledge on  the miscentering parameters from preceding studies. This is a commonly taken approach for many cluster surveys \citep[e.g.][]{McClintock2019,Shin2019,Grandis2024}, where they adopt the miscentering properties calibrated with multi-wavelength data or from simulations as a prior. Here, following \citet{Rykoff2016} and \citet{McClintock2019}, we assume that we have unbiased prior ranges for both miscentering parameters: $f_{\rm mis}=0.22\pm0.11$ and $\sigma_{\rm R}=0.30\pm0.22$. We perform the analysis to the 2D profile data with a flat prior on $\alpha$. 

        The result of fitting the stacked model and the mixture model to our profile measurement data is shown in Fig.~\ref{fig:contour_2D_miscenprior} and the 1-$\sigma$ best-fit ranges for the parameters are shown in the caption. The truth values are marked with the dashed lines, while the posterior for the mixture model (stacked model) is represented by green (yellow) contour. 

        We observe that, even in the presence of the informed miscentering priors, the mixture model results in a $\sim$3 times tighter constraints on the miscentering parameters. 

\section{Summary and discussions}
\label{sec:conclusion}

    Uncertainty in the determination of galaxy cluster centers can be an important systematic in many applications of cluster profiles. In this study, we propose and test a Gaussian mixture model to account for halo miscentering in clusters. The model builds the full likelihood by multiplying the likelihoods for each cluster, thus accommodating cluster-to-cluster variation.

    Using halos within the mass range of $M_{\rm 200m}=[10^{14},10^{14.5}]$ from the 300${\rm Mpc}$ box of IllustrisTNG simulations, we test the performance of the Gaussian mixture model in determining miscentering parameters and compare it with the stacked model that is currently used in the literature.  
    We use the subhalo density profiles around the selected halos as a proxy for profiles measured from observational surveys. We implement miscentering with two parameters: the fraction of miscentered halos ($f_{\rm mis}$) and the amplitude of the miscentering ($\sigma_{\rm R}$) drawn from a Rayleigh distribution. We use two choices for each parameter  ($f_{\rm mis}\in\{0.22,0.50\}$, $\sigma_{\rm R}\in\{0.30,0.50\}$) for a total of four combinations. We then perform model fitting to the subhalo density profiles measured  using the halo model proposed in \citet{Diemer2023}. We work in 3D as well as in 2D (to mimic observational analyses) per miscentering configuration. 

    Our results show that, while both models deliver the posteriors on the miscentering parameters which reasonably agree with the true values, the Gaussian mixture model exhibits significantly improved precision in all cases compared to the stacked model (Figures 4-6). Even when there is prior knowledge on the true miscentering properties, the mixture model still gives several times tighter constraints on the miscentering parameters. The likely reason is that there is sufficient signal-to-noise (S/N) per cluster to get information from the density profile on miscentering, in particular whether the cluster is miscentered or not. The stacked model loses this information on individual clusters.  In the low S/N regime, the expected gain will be lower with the two models converging in the limit of a large number of clusters with very low S/N per cluster. 

    The improved precision achieved by the mixture model motivates its use for observational applications, especially in the small sample regime. There are several open questions and caveats. Our analysis is for subhalo profiles around a sample of over 200 clusters from the IllustrisTNG simulations. Testing it with a bigger cluster sample and for weak lensing or SZ profiles would be a useful test of the methodology: we expect the mixture model to do well with the improved S/N expected in Stage 4 surveys. In addition, it would be worth estimating the bias when using a plausible but incorrect miscentering model.  

    In application to observations, several sources of uncertainty arise, such as anisotropic halo shapes and projection effects which are not well modeled by miscentering. The simulated clusters in our analysis include these two effects but are missing others such as boost factors involved in lensing data that will require additional tests. In addition, the wide range of sample sizes and statistical noise levels in Stage 3 vs. upcoming Stage 4 cosmological surveys will require additional tests. Nevertheless, given the large gain in precision over the stacked model in this study, there is good reason to expect that the Gaussian mixture model will provide an improved description of cluster miscentering and possibly other cluster-to-cluster variations. 

\begin{acknowledgements}
      We are very grateful to Sam Goldstein for discussions and sharing his cluster analysis code. We thank Susmita Adhikari, Gary Bernstein, Chihway Chang and Sam Charney for helpful discussions. BJ is funded partially  by NASA and the US Department of Energy grant DE-SC0007901. TS is funded by the US Department of Energy grant DE-SC0023387. 
\end{acknowledgements}

%
\bibliographystyle{aanda} 
\bibliography{aanda} 

\begin{thebibliography}{37}
\expandafter\ifx\csname natexlab\endcsname\relax\def\natexlab#1{#1}\fi

\bibitem[{{Adhikari} {et~al.}(2016){Adhikari}, {Dalal}, \& {Clampitt}}]{Adhikari2016}
{Adhikari}, S., {Dalal}, N., \& {Clampitt}, J. 2016, \jcap, 2016, 022

\bibitem[{{Adhikari} {et~al.}(2021){Adhikari}, {Shin}, {Jain}, {Hilton}, {Baxter}, {Chang}, {Wechsler}, {Battaglia}, {Bond}, {Bocquet}, {Choi}, {DeRose}, {Devlin}, {Dunkley}, {Evrard}, {Ferraro}, {Hill}, {Hughes}, {Gallardo}, {Lokken}, {MacInnis}, {Madhavacheril}, {McMahon}, {Nati}, {Newburgh}, {Niemack}, {Page}, {Palmese}, {Partridge}, {Rozo}, {Rykoff}, {Salatino}, {Schillaci}, {Sehgal}, {Sif{\'o}n}, {To}, {Wollack}, {Wu}, {Xu}, {Aguena}, {Allam}, {Amon}, {Annis}, {Avila}, {Bacon}, {Bertin}, {Bhargava}, {Brooks}, {Burke}, {Rosell}, {Kind}, {Carretero}, {Castander}, {Choi}, {Costanzi}, {da Costa}, {Vicente}, {Desai}, {Diehl}, {Doel}, {Everett}, {Ferrero}, {Fert{\'e}}, {Flaugher}, {Fosalba}, {Frieman}, {Garc{\'\i}a-Bellido}, {Gaztanaga}, {Gruen}, {Gruendl}, {Gschwend}, {Gutierrez}, {Hartley}, {Hinton}, {Hollowood}, {Honscheid}, {James}, {Jeltema}, {Kuehn}, {Kuropatkin}, {Lahav}, {Lima}, {Maia}, {Marshall}, {Martini}, {Melchior}, {Menanteau}, {Miquel}, {Morgan}, {L.~C. Ogando}, {Paz-Chinch{\'o}n},
  {Malag{\'o}n}, {Sanchez}, {Santiago}, {Scarpine}, {Serrano}, {Sevilla-Noarbe}, {Smith}, {Soares-Santos}, {Suchyta}, {E.~C. Swanson}, {Varga}, {Wilkinson}, {Zhang}, {Austermann}, {Beall}, {Becker}, {Denison}, {Duff}, {Hilton}, {Hubmayr}, {Ullom}, {Lanen}, {Vale}, {Vale}, \& {Vale}}]{Adhikari2021}
{Adhikari}, S., {Shin}, T.-h., {Jain}, B., {et~al.} 2021, \apj, 923, 37

\bibitem[{{Bleem} {et~al.}(2015){Bleem}, {Stalder}, {de Haan}, {Aird}, {Allen}, {Applegate}, {Ashby}, {Bautz}, {Bayliss}, {Benson}, {Bocquet}, {Brodwin}, {Carlstrom}, {Chang}, {Chiu}, {Cho}, {Clocchiatti}, {Crawford}, {Crites}, {Desai}, {Dietrich}, {Dobbs}, {Foley}, {Forman}, {George}, {Gladders}, {Gonzalez}, {Halverson}, {Hennig}, {Hoekstra}, {Holder}, {Holzapfel}, {Hrubes}, {Jones}, {Keisler}, {Knox}, {Lee}, {Leitch}, {Liu}, {Lueker}, {Luong-Van}, {Mantz}, {Marrone}, {McDonald}, {McMahon}, {Meyer}, {Mocanu}, {Mohr}, {Murray}, {Padin}, {Pryke}, {Reichardt}, {Rest}, {Ruel}, {Ruhl}, {Saliwanchik}, {Saro}, {Sayre}, {Schaffer}, {Schrabback}, {Shirokoff}, {Song}, {Spieler}, {Stanford}, {Staniszewski}, {Stark}, {Story}, {Stubbs}, {Vanderlinde}, {Vieira}, {Vikhlinin}, {Williamson}, {Zahn}, \& {Zenteno}}]{Bleem2015}
{Bleem}, L.~E., {Stalder}, B., {de Haan}, T., {et~al.} 2015, \apjs, 216, 27

\bibitem[{{Bocquet} {et~al.}(2024){Bocquet}, {Grandis}, {Bleem}, {Klein}, {Mohr}, {Schrabback}, {Abbott}, {Ade}, {Aguena}, {Alarcon}, {Allam}, {Allen}, {Alves}, {Amon}, {Anderson}, {Annis}, {Ansarinejad}, {Austermann}, {Avila}, {Bacon}, {Bayliss}, {Beall}, {Bechtol}, {Becker}, {Bender}, {Benson}, {Bernstein}, {Bhargava}, {Bianchini}, {Brodwin}, {Brooks}, {Bryant}, {Campos}, {Canning}, {Carlstrom}, {Carnero Rosell}, {Carrasco Kind}, {Carretero}, {Castander}, {Cawthon}, {Chang}, {Chang}, {Chaubal}, {Chen}, {Chiang}, {Choi}, {Chou}, {Citron}, {Corbett Moran}, {Cordero}, {Costanzi}, {Crawford}, {Crites}, {da Costa}, {Pereira}, {Davis}, {Davis}, {DeRose}, {Desai}, {de Haan}, {Diehl}, {Dobbs}, {Dodelson}, {Doux}, {Drlica-Wagner}, {Eckert}, {Elvin-Poole}, {Everett}, {Everett}, {Ferrero}, {Fert{\'e}}, {Flores}, {Frieman}, {Gallicchio}, {Garc{\'\i}a-Bellido}, {Gatti}, {George}, {Giannini}, {Gladders}, {Gruen}, {Gruendl}, {Gupta}, {Gutierrez}, {Halverson}, {Harrison}, {Hartley}, {Herner}, {Hinton}, {Holder},
  {Hollowood}, {Holzapfel}, {Honscheid}, {Hrubes}, {Huang}, {Hubmayr}, {Huff}, {Huterer}, {Irwin}, {James}, {Jarvis}, {Khullar}, {Kim}, {Knox}, {Kraft}, {Krause}, {Kuehn}, {Kuropatkin}, {K{\'e}ruzor{\'e}}, {Lahav}, {Lee}, {Leget}, {Li}, {Lin}, {Lowitz}, {MacCrann}, {Mahler}, {Mantz}, {Marshall}, {McCullough}, {McDonald}, {McMahon}, {Mena-Fern{\'a}ndez}, {Menanteau}, {Meyer}, {Miquel}, {Montgomery}, {Myles}, {Natoli}, {Navarro-Alsina}, {Nibarger}, {Noble}, {Novosad}, {Ogando}, {Omori}, {Padin}, {Pandey}, {Paschos}, {Patil}, {Pieres}, {Plazas Malag{\'o}n}, {Porredon}, {Prat}, {Pryke}, {Raveri}, {Reichardt}, {Roberson}, {Rollins}, {Romero}, {Roodman}, {Ruhl}, {Rykoff}, {Saliwanchik}, {Salvati}, {S{\'a}nchez}, {Sanchez}, {Sanchez Cid}, {Saro}, {Schaffer}, {Secco}, {Sevilla-Noarbe}, {Sharon}, {Sheldon}, {Shin}, {Sievers}, {Smecher}, {Smith}, {Somboonpanyakul}, {Sommer}, {Stalder}, {Stark}, {Stephen}, {Strazzullo}, {Suchyta}, {Tarle}, {To}, {Troxel}, {Tucker}, {Tutusaus}, {Varga}, {Veach}, {Vieira}, {Vikhlinin},
  {von der Linden}, {Wang}, {Weaverdyck}, {Weller}, {Whitehorn}, {Wu}, {Yanny}, {Yefremenko}, {Yin}, {Young}, {Zebrowski}, {Zhang}, {Zohren}, \& {Zuntz}}]{Bocquet2024}
{Bocquet}, S., {Grandis}, S., {Bleem}, L.~E., {et~al.} 2024, arXiv e-prints, arXiv:2401.02075

\bibitem[{{Buchner} {et~al.}(2014){Buchner}, {Georgakakis}, {Nandra}, {Hsu}, {Rangel}, {Brightman}, {Merloni}, {Salvato}, {Donley}, \& {Kocevski}}]{Buchner2014}
{Buchner}, J., {Georgakakis}, A., {Nandra}, K., {et~al.} 2014, \aap, 564, A125

\bibitem[{{Bulbul} {et~al.}(2024){Bulbul}, {Liu}, {Kluge}, {Zhang}, {Sanders}, {Bahar}, {Ghirardini}, {Artis}, {Seppi}, {Garrel}, {Ramos-Ceja}, {Comparat}, {Balzer}, {B{\"o}ckmann}, {Br{\"u}ggen}, {Clerc}, {Dennerl}, {Dolag}, {Freyberg}, {Grandis}, {Gruen}, {Kleinebreil}, {Krippendorf}, {Lamer}, {Merloni}, {Migkas}, {Nandra}, {Pacaud}, {Predehl}, {Reiprich}, {Schrabback}, {Veronica}, {Weller}, \& {Zelmer}}]{Bulbul2024}
{Bulbul}, E., {Liu}, A., {Kluge}, M., {et~al.} 2024, \aap, 685, A106

\bibitem[{{Costanzi} {et~al.}(2019){Costanzi}, {Rozo}, {Simet}, {Zhang}, {Evrard}, {Mantz}, {Rykoff}, {Jeltema}, {Gruen}, {Allen}, {McClintock}, {Romer}, {von der Linden}, {Farahi}, {DeRose}, {Varga}, {Weller}, {Giles}, {Hollowood}, {Bhargava}, {Bermeo-Hernandez}, {Chen}, {Abbott}, {Abdalla}, {Avila}, {Bechtol}, {Brooks}, {Buckley-Geer}, {Burke}, {Rosell}, {Kind}, {Carretero}, {Crocce}, {Cunha}, {da Costa}, {Davis}, {De Vicente}, {Diehl}, {Dietrich}, {Doel}, {Eifler}, {Estrada}, {Flaugher}, {Fosalba}, {Frieman}, {Garc{\'\i}a-Bellido}, {Gaztanaga}, {Gerdes}, {Giannantonio}, {Gruendl}, {Gschwend}, {Gutierrez}, {Hartley}, {Honscheid}, {Hoyle}, {James}, {Krause}, {Kuehn}, {Kuropatkin}, {Lima}, {Lin}, {Maia}, {March}, {Marshall}, {Martini}, {Menanteau}, {Miller}, {Miquel}, {Mohr}, {Ogando}, {Plazas}, {Roodman}, {Sanchez}, {Scarpine}, {Schindler}, {Schubnell}, {Serrano}, {Sevilla-Noarbe}, {Sheldon}, {Smith}, {Soares-Santos}, {Sobreira}, {Suchyta}, {Swanson}, {Tarle}, {Thomas}, \& {Wechsler}}]{Costanzi2019}
{Costanzi}, M., {Rozo}, E., {Simet}, M., {et~al.} 2019, \mnras, 488, 4779

\bibitem[{{Diemer}(2023)}]{Diemer2023}
{Diemer}, B. 2023, \mnras, 519, 3292

\bibitem[{{Dolag} {et~al.}(2009){Dolag}, {Borgani}, {Murante}, \& {Springel}}]{Dolag2009}
{Dolag}, K., {Borgani}, S., {Murante}, G., \& {Springel}, V. 2009, \mnras, 399, 497

\bibitem[{{Genel} {et~al.}(2014){Genel}, {Vogelsberger}, {Springel}, {Sijacki}, {Nelson}, {Snyder}, {Rodriguez-Gomez}, {Torrey}, \& {Hernquist}}]{Genel2014}
{Genel}, S., {Vogelsberger}, M., {Springel}, V., {et~al.} 2014, \mnras, 445, 175

\bibitem[{{Ghirardini} {et~al.}(2024){Ghirardini}, {Bulbul}, {Artis}, {Clerc}, {Garrel}, {Grandis}, {Kluge}, {Liu}, {Bahar}, {Balzer}, {Chiu}, {Comparat}, {Gruen}, {Kleinebreil}, {Krippendorf}, {Merloni}, {Nandra}, {Okabe}, {Pacaud}, {Predehl}, {Ramos-Ceja}, {Reiprich}, {Sanders}, {Schrabback}, {Seppi}, {Zelmer}, {Zhang}, {Bornemann}, {Brunner}, {Burwitz}, {Coutinho}, {Dennerl}, {Freyberg}, {Friedrich}, {Gaida}, {Gueguen}, {Haberl}, {Kink}, {Lamer}, {Li}, {Liu}, {Maitra}, {Meidinger}, {Mueller}, {Miyatake}, {Miyazaki}, {Robrade}, {Schwope}, \& {Stewart}}]{Ghirardini2024}
{Ghirardini}, V., {Bulbul}, E., {Artis}, E., {et~al.} 2024, arXiv e-prints, arXiv:2402.08458

\bibitem[{{Grandis} {et~al.}(2024){Grandis}, {Ghirardini}, {Bocquet}, {Garrel}, {Mohr}, {Liu}, {Kluge}, {Kimmig}, {Reiprich}, {Alarcon}, {Amon}, {Artis}, {Bahar}, {Balzer}, {Bechtol}, {Becker}, {Bernstein}, {Bulbul}, {Campos}, {Carnero Rosell}, {Carrasco Kind}, {Cawthon}, {Chang}, {Chen}, {Chiu}, {Choi}, {Clerc}, {Comparat}, {Cordero}, {Davis}, {Derose}, {Diehl}, {Dodelson}, {Doux}, {Drlica-Wagner}, {Eckert}, {Elvin-Poole}, {Everett}, {Ferte}, {Gatt}, {Giannini}, {Giles}, {Gruen}, {Gruendl}, {Harrison}, {Hartley}, {Herner}, {Huf}, {Kleinebreil}, {Kuropatkin}, {Leget}, {Maccrann}, {Mccullough}, {Merloni}, {Myles}, {Nandra}, {Navarro-Alsina}, {Okabe}, {Pacaud}, {Pandey}, {Prat}, {Predehl}, {Ramos}, {Raveri}, {Rollins}, {Roodman}, {Ross}, {Rykoff}, {Sanchez}, {Sanders}, {Schrabback}, {Secco}, {Seppi}, {Sevilla-Noarbe}, {Sheldon}, {Shin}, {Troxel}, {Tutusaus}, {Varga}, {Wu}, {Yanny}, {Yin}, {Zhang}, {Zhang}, {Alves}, {Bhargava}, {Brooks}, {Burke}, {Carretero}, {Costanzi}, {da Costa}, {Pereira}, {De Vicente},
  {Desai}, {Doel}, {Ferrero}, {Flaugher}, {Friedel}, {Frieman}, {Garc{\'\i}a-Bellido}, {Gutierrez}, {Hinton}, {Hollowood}, {Honscheid}, {James}, {Jeffrey}, {Lahav}, {Lee}, {Marshall}, {Menanteau}, {Ogando}, {Pieres}, {Plazas Malag{\'o}n}, {Romer}, {Sanchez}, {Schubnell}, {Smith}, {Suchyta}, {Swanson}, {Tarle}, {Weaverdyck}, \& {Weller}}]{Grandis2024}
{Grandis}, S., {Ghirardini}, V., {Bocquet}, S., {et~al.} 2024, arXiv e-prints, arXiv:2402.08455

\bibitem[{{Hilton} {et~al.}(2021){Hilton}, {Sif{\'o}n}, {Naess}, {Madhavacheril}, {Oguri}, {Rozo}, {Rykoff}, {Abbott}, {Adhikari}, {Aguena}, {Aiola}, {Allam}, {Amodeo}, {Amon}, {Annis}, {Ansarinejad}, {Aros-Bunster}, {Austermann}, {Avila}, {Bacon}, {Battaglia}, {Beall}, {Becker}, {Bernstein}, {Bertin}, {Bhandarkar}, {Bhargava}, {Bond}, {Brooks}, {Burke}, {Calabrese}, {Carrasco Kind}, {Carretero}, {Choi}, {Choi}, {Conselice}, {da Costa}, {Costanzi}, {Crichton}, {Crowley}, {D{\"u}nner}, {Denison}, {Devlin}, {Dicker}, {Diehl}, {Dietrich}, {Doel}, {Duff}, {Duivenvoorden}, {Dunkley}, {Everett}, {Ferraro}, {Ferrero}, {Fert{\'e}}, {Flaugher}, {Frieman}, {Gallardo}, {Garc{\'\i}a-Bellido}, {Gaztanaga}, {Gerdes}, {Giles}, {Golec}, {Gralla}, {Grandis}, {Gruen}, {Gruendl}, {Gschwend}, {Gutierrez}, {Han}, {Hartley}, {Hasselfield}, {Hill}, {Hilton}, {Hincks}, {Hinton}, {Ho}, {Honscheid}, {Hoyle}, {Hubmayr}, {Huffenberger}, {Hughes}, {Jaelani}, {Jain}, {James}, {Jeltema}, {Kent}, {Knowles}, {Koopman}, {Kuehn}, {Lahav},
  {Lima}, {Lin}, {Lokken}, {Loubser}, {MacCrann}, {Maia}, {Marriage}, {Martin}, {McMahon}, {Melchior}, {Menanteau}, {Miquel}, {Miyatake}, {Moodley}, {Morgan}, {Mroczkowski}, {Nati}, {Newburgh}, {Niemack}, {Nishizawa}, {Ogando}, {Orlowski-Scherer}, {Page}, {Palmese}, {Partridge}, {Paz-Chinch{\'o}n}, {Phakathi}, {Plazas}, {Robertson}, {Romer}, {Carnero Rosell}, {Salatino}, {Sanchez}, {Schaan}, {Schillaci}, {Sehgal}, {Serrano}, {Shin}, {Simon}, {Smith}, {Soares-Santos}, {Spergel}, {Staggs}, {Storer}, {Suchyta}, {Swanson}, {Tarle}, {Thomas}, {To}, {Trac}, {Ullom}, {Vale}, {Van Lanen}, {Vavagiakis}, {De Vicente}, {Wilkinson}, {Wollack}, {Xu}, \& {Zhang}}]{Hilton2021}
{Hilton}, M., {Sif{\'o}n}, C., {Naess}, S., {et~al.} 2021, \apjs, 253, 3

\bibitem[{{Hollowood} {et~al.}(2019){Hollowood}, {Jeltema}, {Chen}, {Farahi}, {Evrard}, {Everett}, {Rozo}, {Rykoff}, {Bernstein}, {Bermeo-Hernandez}, {Eiger}, {Giles}, {Israel}, {Michel}, {Noorali}, {Romer}, {Rooney}, \& {Splettstoesser}}]{Hollowood2019}
{Hollowood}, D.~L., {Jeltema}, T., {Chen}, X., {et~al.} 2019, \apjs, 244, 22

\bibitem[{{Kelly} {et~al.}(2023){Kelly}, {Jobel}, {Eiger}, {Abd}, {Jeltema}, {Giles}, {Hollowood}, {Wilkinson}, {Turner}, {Bhargava}, {Everett}, {Farahi}, {Romer}, {Rykoff}, {Wang}, {Bocquet}, {Cross}, {Faridjoo}, {Franco}, {Gardner}, {Kwiecien}, {Laubner}, {McDaniel}, {O'Donnell}, {Sanchez}, {Schmidt}, {Sripada}, {Swart}, {Upsdell}, {Webber}, {Aguena}, {Allam}, {Alves}, {Bacon}, {Brooks}, {Burke}, {Carnero Rosell}, {Carretero}, {Collins}, {Costanzi}, {da Costa}, {Pereira}, {Davis}, {Doel}, {Ferrero}, {Frieman}, {Garc{\'\i}a-Bellido}, {Giannini}, {Gruen}, {Gruendl}, {Hilton}, {Hinton}, {Honscheid}, {James}, {Kuehn}, {Mann}, {Marshall}, {Mena-Fern{\'a}ndez}, {Miller}, {Miquel}, {Myles}, {Palmese}, {Pieres}, {Plazas Malag{\'o}n}, {Rooney}, {Sahlen}, {Sanchez}, {Sanchez Cid}, {Schubnell}, {Sevilla-Noarbe}, {Smith}, {Stott}, {Suchyta}, {Swanson}, {Tarle}, {To}, {Viana}, {Weaverdyck}, \& {Wiseman}}]{Kelly2023}
{Kelly}, P., {Jobel}, J., {Eiger}, O., {et~al.} 2023, arXiv e-prints, arXiv:2310.13207

\bibitem[{{Mantz} {et~al.}(2015){Mantz}, {von der Linden}, {Allen}, {Applegate}, {Kelly}, {Morris}, {Rapetti}, {Schmidt}, {Adhikari}, {Allen}, {Burchat}, {Burke}, {Cataneo}, {Donovan}, {Ebeling}, {Shandera}, \& {Wright}}]{Mantz2015}
{Mantz}, A.~B., {von der Linden}, A., {Allen}, S.~W., {et~al.} 2015, \mnras, 446, 2205

\bibitem[{{Marinacci} {et~al.}(2018){Marinacci}, {Vogelsberger}, {Pakmor}, {Torrey}, {Springel}, {Hernquist}, {Nelson}, {Weinberger}, {Pillepich}, {Naiman}, \& {Genel}}]{Marinacci2018}
{Marinacci}, F., {Vogelsberger}, M., {Pakmor}, R., {et~al.} 2018, \mnras, 480, 5113

\bibitem[{{McClintock} {et~al.}(2019){McClintock}, {Varga}, {Gruen}, {Rozo}, {Rykoff}, {Shin}, {Melchior}, {DeRose}, {Seitz}, {Dietrich}, {Sheldon}, {Zhang}, {von der Linden}, {Jeltema}, {Mantz}, {Romer}, {Allen}, {Becker}, {Bermeo}, {Bhargava}, {Costanzi}, {Everett}, {Farahi}, {Hamaus}, {Hartley}, {Hollowood}, {Hoyle}, {Israel}, {Li}, {MacCrann}, {Morris}, {Palmese}, {Plazas}, {Pollina}, {Rau}, {Simet}, {Soares-Santos}, {Troxel}, {Vergara Cervantes}, {Wechsler}, {Zuntz}, {Abbott}, {Abdalla}, {Allam}, {Annis}, {Avila}, {Bridle}, {Brooks}, {Burke}, {Carnero Rosell}, {Carrasco Kind}, {Carretero}, {Castander}, {Crocce}, {Cunha}, {D'Andrea}, {da Costa}, {Davis}, {De Vicente}, {Diehl}, {Doel}, {Drlica-Wagner}, {Evrard}, {Flaugher}, {Fosalba}, {Frieman}, {Garc{\'\i}a-Bellido}, {Gaztanaga}, {Gerdes}, {Giannantonio}, {Gruendl}, {Gutierrez}, {Honscheid}, {James}, {Kirk}, {Krause}, {Kuehn}, {Lahav}, {Li}, {Lima}, {March}, {Marshall}, {Menanteau}, {Miquel}, {Mohr}, {Nord}, {Ogando}, {Roodman}, {Sanchez}, {Scarpine},
  {Schindler}, {Sevilla-Noarbe}, {Smith}, {Smith}, {Sobreira}, {Suchyta}, {Swanson}, {Tarle}, {Tucker}, {Vikram}, {Walker}, {Weller}, \& {DES Collaboration}}]{McClintock2019}
{McClintock}, T., {Varga}, T.~N., {Gruen}, D., {et~al.} 2019, \mnras, 482, 1352

\bibitem[{{Melchior} {et~al.}(2017){Melchior}, {Gruen}, {McClintock}, {Varga}, {Sheldon}, {Rozo}, {Amara}, {Becker}, {Benson}, {Bermeo}, {Bridle}, {Clampitt}, {Dietrich}, {Hartley}, {Hollowood}, {Jain}, {Jarvis}, {Jeltema}, {Kacprzak}, {MacCrann}, {Rykoff}, {Saro}, {Suchyta}, {Troxel}, {Zuntz}, {Bonnett}, {Plazas}, {Abbott}, {Abdalla}, {Annis}, {Benoit-L{\'e}vy}, {Bernstein}, {Bertin}, {Brooks}, {Buckley-Geer}, {Carnero Rosell}, {Carrasco Kind}, {Carretero}, {Cunha}, {D'Andrea}, {da Costa}, {Desai}, {Eifler}, {Flaugher}, {Fosalba}, {Garc{\'\i}a-Bellido}, {Gaztanaga}, {Gerdes}, {Gruendl}, {Gschwend}, {Gutierrez}, {Honscheid}, {James}, {Kirk}, {Krause}, {Kuehn}, {Kuropatkin}, {Lahav}, {Lima}, {Maia}, {March}, {Martini}, {Menanteau}, {Miller}, {Miquel}, {Mohr}, {Nichol}, {Ogando}, {Romer}, {Sanchez}, {Scarpine}, {Sevilla-Noarbe}, {Smith}, {Soares-Santos}, {Sobreira}, {Swanson}, {Tarle}, {Thomas}, {Walker}, {Weller}, \& {Zhang}}]{Melchior2017}
{Melchior}, P., {Gruen}, D., {McClintock}, T., {et~al.} 2017, \mnras, 469, 4899

\bibitem[{{Naiman} {et~al.}(2018){Naiman}, {Pillepich}, {Springel}, {Ramirez-Ruiz}, {Torrey}, {Vogelsberger}, {Pakmor}, {Nelson}, {Marinacci}, {Hernquist}, {Weinberger}, \& {Genel}}]{Naiman2018}
{Naiman}, J.~P., {Pillepich}, A., {Springel}, V., {et~al.} 2018, \mnras, 477, 1206

\bibitem[{{Nelson} {et~al.}(2019){Nelson}, {Springel}, {Pillepich}, {Rodriguez-Gomez}, {Torrey}, {Genel}, {Vogelsberger}, {Pakmor}, {Marinacci}, {Weinberger}, {Kelley}, {Lovell}, {Diemer}, \& {Hernquist}}]{Nelson2019}
{Nelson}, D., {Springel}, V., {Pillepich}, A., {et~al.} 2019, Computational Astrophysics and Cosmology, 6, 2

\bibitem[{{Peirani} {et~al.}(2017){Peirani}, {Dubois}, {Volonteri}, {Devriendt}, {Bundy}, {Silk}, {Pichon}, {Kaviraj}, {Gavazzi}, \& {Habouzit}}]{Peirani2017}
{Peirani}, S., {Dubois}, Y., {Volonteri}, M., {et~al.} 2017, \mnras, 472, 2153

\bibitem[{{Pillepich} {et~al.}(2018){Pillepich}, {Nelson}, {Hernquist}, {Springel}, {Pakmor}, {Torrey}, {Weinberger}, {Genel}, {Naiman}, {Marinacci}, \& {Vogelsberger}}]{Pillepich2018}
{Pillepich}, A., {Nelson}, D., {Hernquist}, L., {et~al.} 2018, \mnras, 475, 648

\bibitem[{{Roche} {et~al.}(2024){Roche}, {McDonald}, {Borrow}, {Vogelsberger}, {Shen}, {Springel}, {Hernquist}, {Pakmor}, {Bose}, \& {Kannan}}]{Roche2024}
{Roche}, C., {McDonald}, M., {Borrow}, J., {et~al.} 2024, arXiv e-prints, arXiv:2402.00928

\bibitem[{{Rykoff} {et~al.}(2014){Rykoff}, {Rozo}, {Busha}, {Cunha}, {Finoguenov}, {Evrard}, {Hao}, {Koester}, {Leauthaud}, {Nord}, {Pierre}, {Reddick}, {Sadibekova}, {Sheldon}, \& {Wechsler}}]{Rykoff2014}
{Rykoff}, E.~S., {Rozo}, E., {Busha}, M.~T., {et~al.} 2014, \apj, 785, 104

\bibitem[{{Rykoff} {et~al.}(2016){Rykoff}, {Rozo}, {Hollowood}, {Bermeo-Hernandez}, {Jeltema}, {Mayers}, {Romer}, {Rooney}, {Saro}, {Vergara Cervantes}, {Wechsler}, {Wilcox}, {Abbott}, {Abdalla}, {Allam}, {Annis}, {Benoit-L{\'e}vy}, {Bernstein}, {Bertin}, {Brooks}, {Burke}, {Capozzi}, {Carnero Rosell}, {Carrasco Kind}, {Castander}, {Childress}, {Collins}, {Cunha}, {D'Andrea}, {da Costa}, {Davis}, {Desai}, {Diehl}, {Dietrich}, {Doel}, {Evrard}, {Finley}, {Flaugher}, {Fosalba}, {Frieman}, {Glazebrook}, {Goldstein}, {Gruen}, {Gruendl}, {Gutierrez}, {Hilton}, {Honscheid}, {Hoyle}, {James}, {Kay}, {Kuehn}, {Kuropatkin}, {Lahav}, {Lewis}, {Lidman}, {Lima}, {Maia}, {Mann}, {Marshall}, {Martini}, {Melchior}, {Miller}, {Miquel}, {Mohr}, {Nichol}, {Nord}, {Ogando}, {Plazas}, {Reil}, {Sahl{\'e}n}, {Sanchez}, {Santiago}, {Scarpine}, {Schubnell}, {Sevilla-Noarbe}, {Smith}, {Soares-Santos}, {Sobreira}, {Stott}, {Suchyta}, {Swanson}, {Tarle}, {Thomas}, {Tucker}, {Uddin}, {Viana}, {Vikram}, {Walker}, {Zhang}, \& {DES
  Collaboration}}]{Rykoff2016}
{Rykoff}, E.~S., {Rozo}, E., {Hollowood}, D., {et~al.} 2016, \apjs, 224, 1

\bibitem[{{Seppi} {et~al.}(2023){Seppi}, {Comparat}, {Nandra}, {Dolag}, {Biffi}, {Bulbul}, {Liu}, {Ghirardini}, \& {Ider-Chitham}}]{Seppi2023}
{Seppi}, R., {Comparat}, J., {Nandra}, K., {et~al.} 2023, \aap, 671, A57

\bibitem[{{Shin} {et~al.}(2019){Shin}, {Adhikari}, {Baxter}, {Chang}, {Jain}, {Battaglia}, {Bleem}, {Bocquet}, {DeRose}, {Gruen}, {Hilton}, {Kravtsov}, {McClintock}, {Rozo}, {Rykoff}, {Varga}, {Wechsler}, {Wu}, {Zhang}, {Aiola}, {Allam}, {Bechtol}, {Benson}, {Bertin}, {Bond}, {Brodwin}, {Brooks}, {Buckley-Geer}, {Burke}, {Carlstrom}, {Carnero Rosell}, {Carrasco Kind}, {Carretero}, {Castander}, {Choi}, {Cunha}, {Crawford}, {da Costa}, {De Vicente}, {Desai}, {Devlin}, {Dietrich}, {Doel}, {Dunkley}, {Eifler}, {Evrard}, {Flaugher}, {Fosalba}, {Gallardo}, {Garc{\'\i}a-Bellido}, {Gaztanaga}, {Gerdes}, {Gralla}, {Gruendl}, {Gschwend}, {Gupta}, {Gutierrez}, {Hartley}, {Hill}, {Ho}, {Hollowood}, {Honscheid}, {Hoyle}, {Huffenberger}, {Hughes}, {James}, {Jeltema}, {Kim}, {Krause}, {Kuehn}, {Lahav}, {Lima}, {Madhavacheril}, {Maia}, {Marshall}, {Maurin}, {McMahon}, {Menanteau}, {Miller}, {Miquel}, {Mohr}, {Naess}, {Nati}, {Newburgh}, {Niemack}, {Ogando}, {Page}, {Partridge}, {Patil}, {Plazas}, {Rapetti}, {Reichardt},
  {Romer}, {Sanchez}, {Scarpine}, {Schindler}, {Serrano}, {Smith}, {Smith}, {Soares-Santos}, {Sobreira}, {Staggs}, {Stark}, {Stein}, {Suchyta}, {Swanson}, {Tarle}, {Thomas}, {van Engelen}, {Wollack}, \& {Xu}}]{Shin2019}
{Shin}, T., {Adhikari}, S., {Baxter}, E.~J., {et~al.} 2019, \mnras, 487, 2900

\bibitem[{{Shin} {et~al.}(2021){Shin}, {Jain}, {Adhikari}, {Baxter}, {Chang}, {Pandey}, {Salcedo}, {Weinberg}, {Amsellem}, {Battaglia}, {Belyakov}, {Dacunha}, {Goldstein}, {Kravtsov}, {Varga}, {Abbott}, {Aguena}, {Alarcon}, {Allam}, {Amon}, {Andrade-Oliveira}, {Annis}, {Bacon}, {Bechtol}, {Becker}, {Bernstein}, {Bertin}, {Bocquet}, {Bond}, {Brooks}, {Buckley-Geer}, {Burke}, {Campos}, {Rosell}, {Kind}, {Carretero}, {Chen}, {Choi}, {Costanzi}, {da Costa}, {DeRose}, {Desai}, {De Vicente}, {Devlin}, {Diehl}, {Dietrich}, {Dodelson}, {Doel}, {Doux}, {Drlica-Wagner}, {Eckert}, {Elvin-Poole}, {Everett}, {Ferraro}, {Ferrero}, {Fert{\'e}}, {Flaugher}, {Frieman}, {Gallardo}, {Gatti}, {Gaztanaga}, {Gerdes}, {Gruen}, {Gruendl}, {Gutierrez}, {Harrison}, {Hartley}, {Hill}, {Hilton}, {Hinton}, {Hollowood}, {Hughes}, {James}, {Jarvis}, {Jeltema}, {Koopman}, {Krause}, {Kuehn}, {Kuropatkin}, {Lahav}, {Lima}, {Lokken}, {MacCrann}, {Madhavacheril}, {Maia}, {McCullough}, {McMahon}, {Melchior}, {Menanteau}, {Miquel}, {Mohr},
  {Moodley}, {Morgan}, {Myles}, {Nati}, {Navarro-Alsina}, {Niemack}, {Ogando}, {Page}, {Palmese}, {Partridge}, {Paz-Chinch{\'o}n}, {Pereira}, {Pieres}, {Malag{\'o}n}, {Prat}, {Raveri}, {Rodriguez-Monroy}, {Rollins}, {Romer}, {Rykoff}, {Salatino}, {S{\'a}nchez}, {Sanchez}, {Santiago}, {Scarpine}, {Schillaci}, {Secco}, {Serrano}, {Sevilla-Noarbe}, {Sheldon}, {Sherwin}, {Sif{\'o}n}, {Smith}, {Soares-Santos}, {Staggs}, {Suchyta}, {Swanson}, {Tarle}, {Thomas}, {To}, {Troxel}, {Tutusaus}, {Vavagiakis}, {Weller}, {Wollack}, {Yanny}, {Yin}, \& {Zhang}}]{Shin2021}
{Shin}, T., {Jain}, B., {Adhikari}, S., {et~al.} 2021, \mnras, 507, 5758

\bibitem[{{Sijacki} {et~al.}(2015){Sijacki}, {Vogelsberger}, {Genel}, {Springel}, {Torrey}, {Snyder}, {Nelson}, \& {Hernquist}}]{Sijacki2015}
{Sijacki}, D., {Vogelsberger}, M., {Genel}, S., {et~al.} 2015, \mnras, 452, 575

\bibitem[{{Springel} {et~al.}(2018){Springel}, {Pakmor}, {Pillepich}, {Weinberger}, {Nelson}, {Hernquist}, {Vogelsberger}, {Genel}, {Torrey}, {Marinacci}, \& {Naiman}}]{Springel2018}
{Springel}, V., {Pakmor}, R., {Pillepich}, A., {et~al.} 2018, \mnras, 475, 676

\bibitem[{{Springel} {et~al.}(2001){Springel}, {White}, {Tormen}, \& {Kauffmann}}]{Springel2001}
{Springel}, V., {White}, S. D.~M., {Tormen}, G., \& {Kauffmann}, G. 2001, \mnras, 328, 726

\bibitem[{{Sunyaev} \& {Zeldovich}(1972)}]{SZ1972}
{Sunyaev}, R.~A. \& {Zeldovich}, Y.~B. 1972, Comments on Astrophysics and Space Physics, 4, 173

\bibitem[{{Vogelsberger} {et~al.}(2014{\natexlab{a}}){Vogelsberger}, {Genel}, {Springel}, {Torrey}, {Sijacki}, {Xu}, {Snyder}, {Bird}, {Nelson}, \& {Hernquist}}]{Vogelsberger2014b}
{Vogelsberger}, M., {Genel}, S., {Springel}, V., {et~al.} 2014{\natexlab{a}}, \nat, 509, 177

\bibitem[{{Vogelsberger} {et~al.}(2014{\natexlab{b}}){Vogelsberger}, {Genel}, {Springel}, {Torrey}, {Sijacki}, {Xu}, {Snyder}, {Nelson}, \& {Hernquist}}]{Vogelsberger2014a}
{Vogelsberger}, M., {Genel}, S., {Springel}, V., {et~al.} 2014{\natexlab{b}}, \mnras, 444, 1518

\bibitem[{{Yan} {et~al.}(2020){Yan}, {Raza}, {Van Waerbeke}, {Mead}, {McCarthy}, {Tr{\"o}ster}, \& {Hinshaw}}]{Yan2020}
{Yan}, Z., {Raza}, N., {Van Waerbeke}, L., {et~al.} 2020, \mnras, 493, 1120

\bibitem[{{Zhang} {et~al.}(2019){Zhang}, {Jeltema}, {Hollowood}, {Everett}, {Rozo}, {Farahi}, {Bermeo}, {Bhargava}, {Giles}, {Romer}, {Wilkinson}, {Rykoff}, {Mantz}, {Diehl}, {Evrard}, {Stern}, {Gruen}, {von der Linden}, {Splettstoesser}, {Chen}, {Costanzi}, {Allen}, {Collins}, {Hilton}, {Klein}, {Mann}, {Manolopoulou}, {Morris}, {Mayers}, {Sahlen}, {Stott}, {Vergara Cervantes}, {Viana}, {Wechsler}, {Allam}, {Avila}, {Bechtol}, {Bertin}, {Brooks}, {Burke}, {Carnero Rosell}, {Carrasco Kind}, {Carretero}, {Castander}, {da Costa}, {De Vicente}, {Desai}, {Dietrich}, {Doel}, {Flaugher}, {Fosalba}, {Frieman}, {Garc{\'\i}a-Bellido}, {Gaztanaga}, {Gruendl}, {Gschwend}, {Gutierrez}, {Hartley}, {Honscheid}, {Hoyle}, {Krause}, {Kuehn}, {Kuropatkin}, {Lima}, {Maia}, {Marshall}, {Melchior}, {Menanteau}, {Miller}, {Miquel}, {Ogando}, {Plazas}, {Sanchez}, {Scarpine}, {Schindler}, {Serrano}, {Sevilla-Noarbe}, {Smith}, {Soares-Santos}, {Suchyta}, {Swanson}, {Tarle}, {Thomas}, {Tucker}, {Vikram}, {Wester}, \& {DES
  Collaboration}}]{Zhang2019}
{Zhang}, Y., {Jeltema}, T., {Hollowood}, D.~L., {et~al.} 2019, \mnras, 487, 2578

\end{thebibliography}
%

\end{document}